\documentclass[10pt, letterpaper]{article}
\usepackage[OME]{express}

\usepackage{color}
\usepackage{soul}
\usepackage[normalem]{ulem}
\usepackage{cancel}

\usepackage{latexsym, amssymb, amsfonts, euscript, amsmath}
\usepackage{multirow}

\newcommand{\pdet}{p_{\!_{_\mathrm{D}}}}
\newcommand{\pfr}{p_{_\mathrm{fr}}}

\newcommand{\lb}{\left\lgroup}
\newcommand{\rb}{\right\rgroup}
\newcommand{\var}[1]{\lb #1 \rb}

\newcommand{\p}[1]{p_{_#1}}
\begin{document}


\title{Clarification on Generalized Lau condition for X-ray interferometers based on dual phase gratings}

\author{Aimin Yan,\authormark{1} Xizeng Wu,\authormark{1,*} and Hong Liu\authormark{2}}

\address{\authormark{1}Department of Radiology, University of Alabama at Birmingham, Birmingham, AL 35249, USA \\
\authormark{2}Center for Bioengineering and School of Electrical and Computer Engineering, University of Oklahoma, Norman, OK 73019, USA}

\email{\authormark{*}xwu@uabmc.edu} 



\begin{abstract}
To implement dual phase grating x-ray interferometry with x-ray tubes, one needs to incorporate an absorbing source grating. In order to attain good fringe visibility, the period of a source grating should be subject to a stringent condition. In literature some authors claim that the Lau-condition in Talbot-Lau interferometry can be literally transferred to dual phase grating interferometry. In this work we show that this statement in literature is incorrect. Instead, through an intuitive geometrical analysis of fringe formation, we derived a new generalized Lau-condition that provides a useful design tool for implementation of dual phase grating interferometry.
\end{abstract}






\section{Introduction}\label{sec-intr}
\begin{figure}[htbp]
\centering\includegraphics[width=\textwidth]{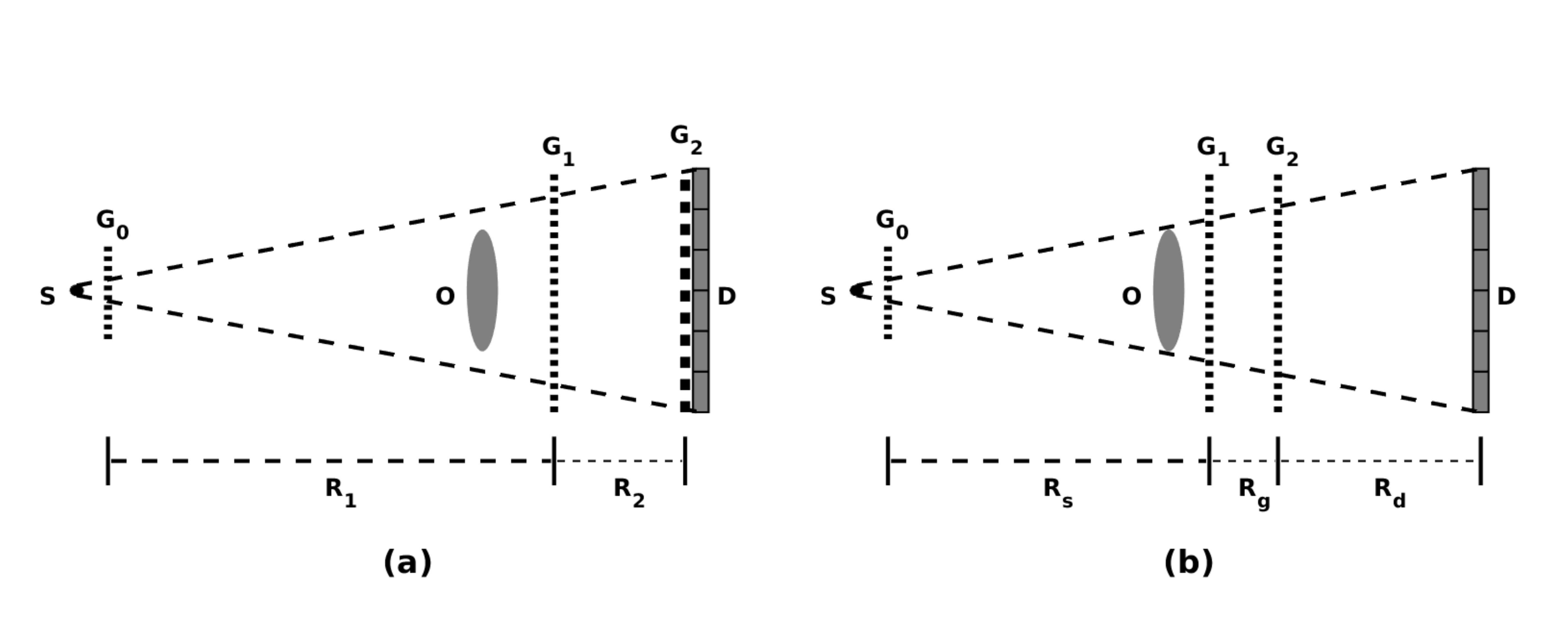}
\caption{Schematic of Talbot-Lau interferometry (a), and dual-phase grating interferometry (b) with source gratings. } \label{fig-setup}
\end{figure}

Currently, the Talbot-Lau x-ray interferometry is widely used for x-ray differential phase contrast imaging\cite{Momose, Weitkamp, Momose-2, Pfeiffer, Yashiro, Zhu-Zhang, Bevins, Tang-Yang-2, Teshima, Bennett, Jiang-Wyatt,Itoh,Suortti, Momose-Huwakara, Morimoto-Fujino, Morimoto, Y-W-L, Y-W-L-2, Y-W-L-3, Y-W-L-4, Donath-Pfeiffer}. In Talbot-Lau grating-based interferometry, a phase grating is employed as a beam splitter to split x-ray into diffraction orders. The interference between the diffracted orders forms intensity fringes. The sample imprints x-ray beam with phase shifts and distorts the intensity fringes. Analyzing the fringes one reconstructs  sample attenuation, phase gradient and dark field images\cite{Momose, Weitkamp, Momose-2, Pfeiffer, Yashiro, Zhu-Zhang, Bevins, Tang-Yang-2, Teshima, Bennett}. To increase the grating interferometer's sensitivity, one needs to use fine-pitch phase gratings of periods as small as few micrometers. But in medical imaging and material science applications, it is only feasible to utilize common imaging detectors, whose pixels are of a few tens of micrometers.  To enable fringe detection with common image detectors, one way is to use a fine absorbing grating placed as shown  in Fig.~\ref{fig-setup}(a) to indirectly detect fringe patterns through grating scanning, which is also called phase stepping procedure\cite{Momose, Weitkamp, Momose-2, Pfeiffer}. However, the absorbing grating blocks more than half of transmitting x-ray, and will increase radiation dose  in imaging exams.  

Recently, dual phase grating x-ray interferometry demonstrated its attractive advantages\cite{Miao, Kagias}.   A typical dual phase grating interferometer employs two phase gratings $G_1$ and $G_2$ as the beam splitters, as is shown in Fig.~\ref{fig-setup}(b).  The split waves transmitting through the phase gratings interfere with each other, creating different diffraction orders, including a beat pattern in the intensity fringe  pattern\cite{Y-W-L-dual-grating-1}.  The imaging detector $D$, as shown in Fig.~\ref{fig-setup}(b), has a pixel size much larger than periods of both the phase gratings.   Due to pixel averaging effects, the detector  just resolves the beat patterns of large periodicities.  In this way, a common imaging detector with resolution of tens micrometers can directly resolve the fringes generated by fine phase gratings. Hence, different from Talbot-Lau setups\cite{Momose, Weitkamp, Momose-2, Pfeiffer, Yashiro, Zhu-Zhang, Bevins, Tang-Yang-2, Teshima, Bennett},  dual phase grating enables direct fringe detection without need of absorbing analyzer grating. This advantage brings significant radiation dose reduction as compared to Talbot-Lau interferometry. Moreover, different from the inverse geometry setups of Talbot-Lau interferometry\cite{Donath-Pfeiffer}, a  dual phase grating  interferometer  keeps the system's length compact, since the fringe period can be conveniently tuned by adjusting the grating-spacing, without increase of system size\cite{Miao, Kagias, Y-W-L-dual-grating-1}.

\begin{figure}[htbp]
\centering\includegraphics[width=\textwidth]{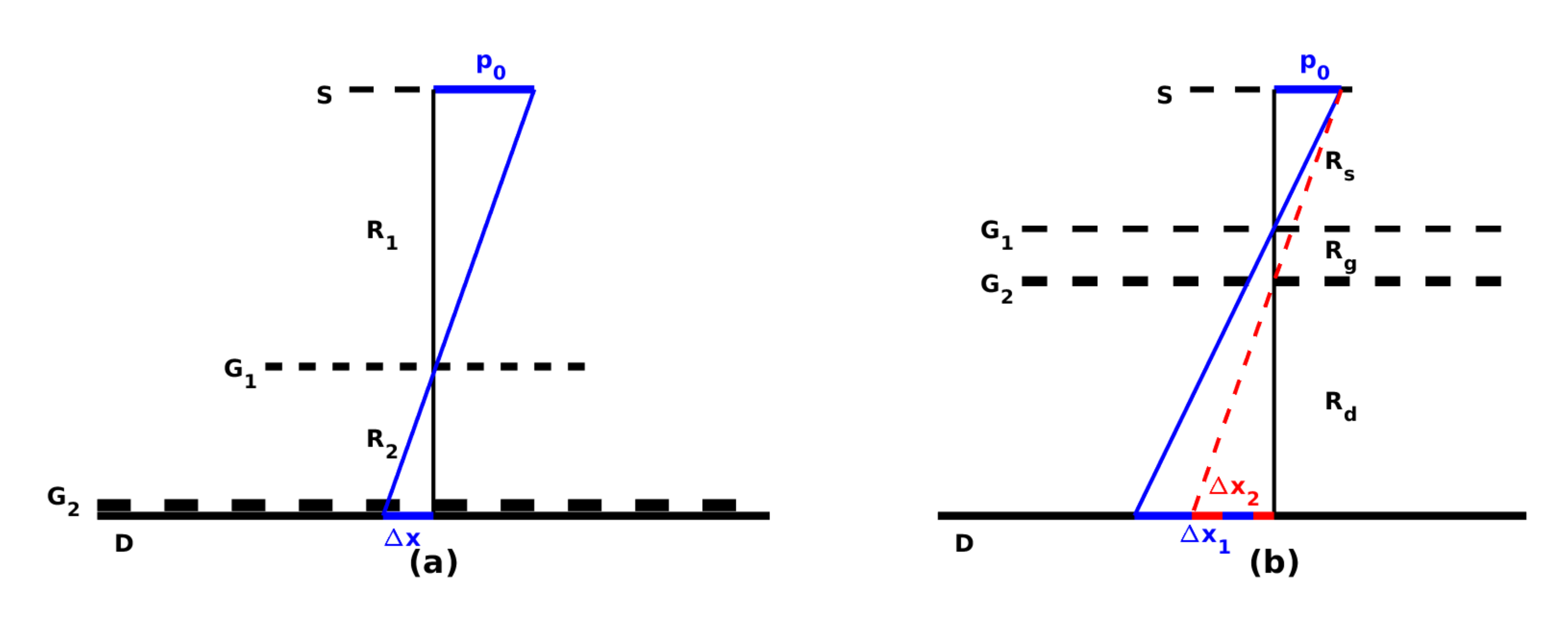}
\caption{Geometric configuration of source grating period $p_0$ of Talbot-Lau interferometry (a) and dual-phase grating interferometry (b). } \label{fig-source-geometry}
\end{figure}

In order to attain good fringe visibility in grating-based x-ray interferometry, a necessary condition is to provide spatially coherent illumination of phase gratings, such as that achieved by use of synchrotron radiation or micro-focus x-ray tubes. But for potential medical imaging applications of dual grating interferometry, it is challenging to find robust x-ray sources to provide sufficient spatial coherence and adequate x-ray flux.  As a solution, one may employ an x-ray tube equipped
with an absorbing source grating $G_0$, which serves as an aperture mask to divide the focus spot into an array of mutually incoherent virtual line sources\cite{Pfeiffer, Momose-Huwakara}. Similarly, one may utilize
a periodic array $G_0$ of micro-anode-sources embedded in the anode for the same purpose\cite{Morimoto-Fujino, Morimoto}.  For Talbot-Lau interferometry with a source grating, the well-known Lau condition on the period of a source grating should be satisfied for  attaining good fringe visibility\cite{Pfeiffer, Momose-Huwakara}. The Lau condition for incorporating a source grating has an intuitive geometric interpretation. When Lau condition is satisfied, each displayed line-source of a source grating should shift the fringe pattern by an integral multiples of the fringe period. With this condition satisfied, the fringe pattern generated by all the line-sources in the source grating are constructively superimposed\cite{Pfeiffer, Momose-Huwakara, Morimoto-Fujino, Morimoto}. Figure~\ref{fig-source-geometry}(a) shows a schematic of geometric configuration of a typical Talbot-Lau interferometer. It is easy to see from Fig.~\ref{fig-source-geometry}(a) that the fringe generated by this line source is displaced by $\Delta x=(R_2/R_1)\p0$, where $\p0$ is the pitch of the source grating, $R_1$ is source-grating $G_0$-phase grating $G_1$ distance, and $R_2$ is the distance between $G_1$-grating and  the analyzer grating $G_2$, which is placed in front of the detector. For constructive superposition one requires $\Delta x$ should be equal to fringe period $\pfr$. In Talbot-Lau interferometry the fringe period $\pfr$ is generally equal to the projection $p_{\mathrm{proj}}$ of the period of phase grating $G_1$, that is,  $\pfr=p_{\mathrm{proj}}=(R_1+R_2)/R_1\p1$. Note that for $\pi$-grating, the fringe period is reduced to $\pfr=p_{\mathrm{proj}}/2$. But with polychromatic x-ray, phase shift of a grating varies with photon energy. Hence, the fringe period of a Talbot-Lau interferometry with polychromatic x-ray is equal to the projection $p_{\mathrm{proj}}$. Note that in Talbot-lau interferometry the period $\p2$ of the absorbing grating is always set to $\p2=\pfr$. Since $\Delta x= \pfr$ is required for coherent superposition of fringes, so  the source grating pitch should be set to 
\begin{equation}\label{eq-lau}
\p0=\frac{R_1}{R_2}\p2.
\end{equation}
This is the so-called Lau condition for constructive fringe superposition in Talbot-Lau interferometry.

But for dual phase grating interferometry, the fringes are generated by two phase gratings, so the the original Lau condition of Eq.~(\ref{eq-lau}) is not applicable anymore. In order to implement the dual phase grating interferometry for wide applications in medical imaging, there is a pressing need to generalize the original Lau condition to dual phase grating interferometry.  In literature a recently published paper  presented an analysis on the validity of the Lau condition  for dual phase grating interferometry\cite{Bopp}. In that paper the authors postulated that the intensity fringe pattern results from low frequency Moire pattern generated by $G_1$ and $G_2$ at the $G_2$-plane\cite{Bopp}. Under this hypothesis, the paper claims that the original Lau condition on the source grating is literally transferred to dual phase grating interferometry.  Rewriting Eq.~(\ref{eq-lau}) with the notations employed in Fig.~\ref{fig-setup}(b) for a dual phase grating interferometer, the authors of that paper claimed that the source grating pitch should satisfy following  condition for achieving coherent fringe superposition in dual phase grating interferometry: 

\begin{equation}\label{eq-p0*}
\p0^{\ast}=\frac{R_s}{R_g} p_{\mathrm{proj}}=\frac{R_s+R_g}{R_g}\p1.
\end{equation}
Here $R_s$ is the distance from source grating to the first grating $G_1$, and $R_g$ is the spacing between the two phase gratings, as is shown in Fig.~\ref{fig-setup}(b).
For sake of convenience in discussion we denote the source grating period determined by above equation (Eq.~(\ref{eq-p0*})) as $\p0^{\ast}$. Note that this $\p0^{\ast}$ given by Eq.~(\ref{eq-p0*}) is independent of the distance $R_d$ between $G_2$-grating and detector. Unfortunately, the Lau condition of Eq.~(\ref{eq-p0*}) for dual phase grating interferometry is incorrect. It fails to take into account the wave propagation from $G_2$ grating to the detector. Setting out to clarify on this issue, in this work we present an intuitive derivation of the condition of constructive fringe superposition in dual phase grating interferometry with a source grating. In section~\ref{sec-method}, we present an intuitive derivation of the generalized Lau condition for dual phase grating interferometry incorporating a source grating.  In section~\ref{sec-results}, we present simulation results that validate our generalized Lau condition for dual phase grating interferometry.  We conclude the work in section~\ref{sec-conclude}.  We hope that our clarification on the generalized Lau condition may help researchers  in design and implementation of dual phase grating x-ray interferometry incorporating a source grating.

\section{Methods}\label{sec-method}
We consider a typical setup of dual phase x-ray interferometry, as is shown in Fig.~\ref{fig-setup}(b).  Different from conventional Talbot-Lau X-ray interferometry, which uses only a single phase-grating, the new technique employs two phase gratings $G_1$ and $G_2$ as the beam splitters. For high-sensitivity the phase gratings $G_1$ and $G_2$ are made of small periods of close to one micrometer, and we assume that $G_1$'s period $\p1$ is equal or close to $G_2$'s period $\p2$.  In addition, the detector has pixel size $\pdet$ that is about few tens of micrometers, so detector pixel size $\pdet \!\!\gg\!\! \p1, \p2$.  In order to employ sources of large focal spots, a strongly absorbing source grating $G_0$ of narrow slits is used to break source into an array of mutually incoherent virtual line sources. But  in order to have the fringes generated by these line sources constructively superimposed, the period $\p0$ of the source grating should satisfy a certain condition, which we called the generalized Lau condition  in dual phase-grating interferometry.  Our task in this work is to derive the generalized Lau condition by using a simple geometric analysis.

Let us consider a setup with a single narrow-slit source first. For sake of convenience in discussion, we define several geometric magnification factors $M_{g_2}$ and $M_{g_1}$ as:
\begin{equation}\label{eq-Ms}
M_{g_1}=\frac{R_s+R_g+R_d}{R_s}; \qquad \qquad M_{g_2}=\frac{R_s+R_g+R_d}{R_s+R_g},
\end{equation}
where $M_{g_1}$ is the geometric magnification factor from the $G_1$ plane to detector plane, and $M_{g_2}$ is the geometric magnification factor from the $G_2$ plane to detector plane. Our analysis starts from the x-ray irradiance $I_{R_s +R_g +R_d} (x, y)$ at detector entrance.  Obviously, in absence of $G_2$ grating, an intensity  diffraction order generated by $G_1$ phase grating alone would be represented by $\exp\left[i2\pi l x/(M_{g_1}\p1)\right]$, where $l$ is an integer and $\p1$ is the period of $G_1$ grating. Similarly, an intensity diffraction order generated by the second phase grating $G_2$ alone  would be denoted by $\exp\left[i2\pi r x/(M_{g_2}\p2)\right]$, where $r$ is an integer and $\p2$ is the period of the second phase grating.  As is shown from a quantitative theory of dual phase grating interferometry\cite{Y-W-L-dual-grating-1}, x-ray irradiance at detector entrance is a result of cross-modulation between the intensity patterns generated by the two phase-gratings $G_1$ and $G_2$ respectively.  Specifically, the intensity pattern is a weighted sum of different diffracted orders. Each of the diffraction orders in dual phase grating interferometry is represented by a product of 
\begin{equation}
\exp\left[i2\pi l \frac{x}{M_{g_1}\p1}\right]\cdot\exp\left[i2\pi r \frac{x}{M_{g_2}\p2}\right] = \exp\left[i2\pi x\cdot\left(\frac{l}{M_{g_1}\p1} + \frac{r}{M_{g_2}\p2}\right)\right],
\end{equation}
and each of the diffraction orders is indexed by two integers $(l,r)$. It is easy to see from above equation that the intensity fringe of order $(l,r)$ has a periodicity of $\left(l/(M_{g_1}\p1)+r/(M_{g_2}\p2)\right)^{-1}$. Since grating periods are few micrometers only, so the interference fringes are generally so fine that most of them are too fine to be detected by common imaging detector. However, among the fringe patterns, there are beat patterns formed by those diffraction orders characterized by $l=-r$. These beat patterns are indeed generated by the interference beat patterns. The beat pattern consists of fundamental frequency and its harmonics resolved by the imaging detector. The period of the first harmonics of beat patterns is\cite{Y-W-L-dual-grating-1}
\begin{equation}\label{eq-pfr}
\pfr=\var{\frac{-1}{M_{g_1}\p1}+\frac{1}{M_{g_2}\p2}}^{-1} = \frac{R_s+R_g+R_d}{R_g/\p2+R_s(1/\p2-1/\p1)}.
\end{equation}
On the other hand, period of the $l$-th harmonics is $\pfr/l$. Hence the period of the intensity fringe is $\pfr$ of Eq.~(\ref{eq-pfr}). According to Eq.~(\ref{eq-pfr}), the beat pattern periods can be much larger than grating periods $\p1$ and $\p2$. For example, if the system geometry is set in such a way that $R_s+R_d\!\gg\! R_g$ and $\p1\approx \p2$ then the resulting fringe period $\pfr\!\gg\! \p1,\p2$. Consequently, the beat pattern may be resolved by a common imaging detector.  On the other hand, as long as detector period $\pdet\!\gg\! \p1,\p2$, the detector renders all other fine fringes with $l\neq -r$ orders to a constant background, owing to detector-pixel averaging effect.

Once the fringe formation mechanism is understood, we are now ready to derive the generalized Lau condition for a dual phase grating setup incorporating a source grating. Consider an off-center line source that is displaced by $\p0$ from the center. As is demonstrated in Fig.~\ref{fig-source-geometry}(b), from the similar triangles relationship, this source-displacement $\p0$ will cause a fringe shift $\Delta x_1 = \p0(R_g+R_d)/R_s$ for the $G_1$-associated fringes, so its diffraction order will become $\exp\left[i2\pi l(x+\Delta x_1)/(M_{g_1}\p1)\right]$. By the same reasoning, this off-center line source  causes a fringe shift $\Delta x_2 = \p0 R_d/(R_s+R_g)$ for the $G_2$-associated fringes, so its diffraction order will become $\exp\left[i2\pi r(x+\Delta x_2)/(M_{g_2}\p2)\right]$. Since the fringes resolved by the detector are the beat patterns formed by cross-modulation between the $G_1$ and $G_2$-associated fringes with diffracted orders $l=-r$, so this off-center line source generates a total fringe phase shift (in radians) $\Delta\Phi=2\pi l\left(\Delta x_1/(M_{g_1}\p1)-\Delta x_2/(M_{g_2}\p2)\right)$ in the resolved intensity fringe.  To achieve constructive fringe superposition for all diffraction orders, one should make $\Delta x_1/(M_{g_1}\p1) - \Delta x_2/(M_{g_2}\p2)=1$,  or equivalently, one should require:
\begin{equation}
 \var{\frac{R_g+R_d}{M_{g_1} R_s \p1} - \frac{R_d}{M_{g_2}(R_s+R_g)\p2}}\cdot \p0=1.
\end{equation}
Using Eq.~(\ref{eq-Ms}), we can rewrite the condition of fringe constructive superposition as:
\begin{equation}\label{eq-p0}
\p0=\frac{R_s+R_g+R_d}{R_g/\p1 + R_d(1/\p1-1/\p2)}
\end{equation}
Equation~(\ref{eq-p0}) is the generalized Lau condition, for dual phase grating interferometry incorporating a source grating. This condition determines the source grating periodicity for achieving constructive superposition of fringes generated by all the line sources in a source grating. 

\section{Results}\label{sec-results}
In order to validate the generalized Lau condition derived above we conducted  numerical simulations, in which four interferometer setups were simulated. The four setups  employ the same phase gratings and identical geometric configurations, such as $R_s$, $R_g$, and $R_d$, but differ in source configurations.   In the first setup, the interferometer consists of a 20-keV point source,  two $\pi$-phase gratings of periods $\p1=1\mu\mathrm{m}$, and $\p2=1.1\mu\mathrm{m}$ respectively, and a detector of $19.3 \mu$m pixels.  The geometry  configuration of this interferometer  was  set to $R_s=45$cm, $R_g=4$cm, and $R_d=38$cm. This geometric configuration was selected according to the fringe visibility formulas for good fringe visibility\cite{Y-W-L-dual-grating-1}. This setup serves as a reference for fringe visibility comparison, as the point source is spatially coherent.  For the other three interferometer setups, we replaced the point source with a focal spot of 0.35 mm in width, while keeping otherwise the same geometric configuration as the first setup. In the second setup no source grating was used. In the third and fourth setups different source gratings were placed in front of the focal spot respectively. 

\begin{figure}[htbp]
\centering\includegraphics[width=\textwidth]{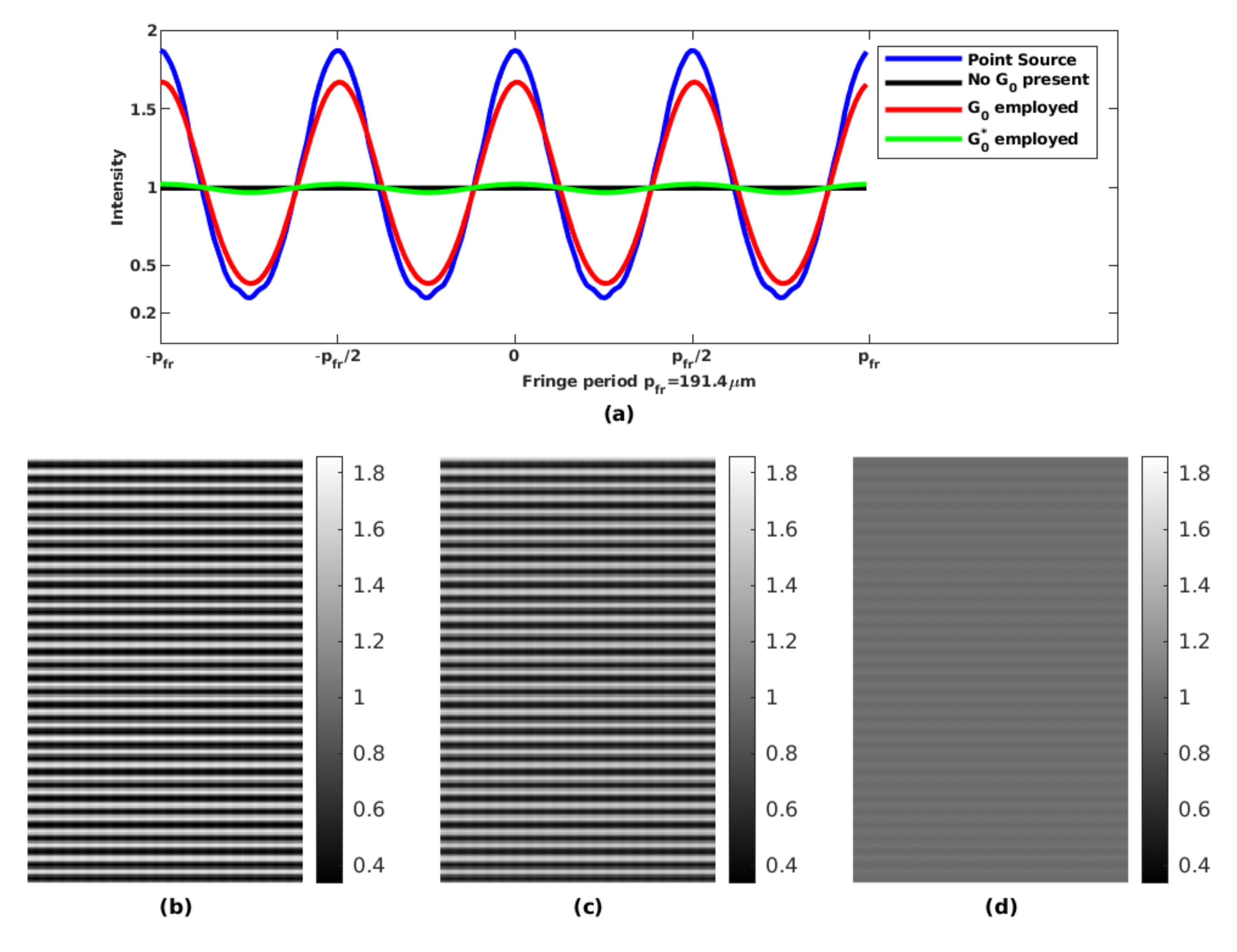}
\caption{Numerical simulation of Talbot-Lau effect in dual-phase grating interferometry.  The four curves in Fig.~\ref{fig-simu-nonEqualP}(a) plot the fringe  intensity values over two fringe periods with  one period  equals to $191.4\mu$m, corresponding to the four different settings. The blue curve is the plot for the setup with an ideal point source, while the red and green curves correspond to the setups in which the source grating's period is set according to Eq.~(\ref{eq-p0}) and Eq.~(\ref{eq-p0*}) respectively. The black curve is the plot for the setup without use of source grating.  The images in Fig.~\ref{fig-simu-nonEqualP}(b), (c) and (d) are the corresponding intensity maps when detector pixel size is set to $19.33\mu$m.  These intensity maps are corresponding to setups associated with the blue, red and green curves in Fig. 3(a). For details, see text.} \label{fig-simu-nonEqualP}
\end{figure}

In the third setup a source grating of  period $p_0=11.67\mu$m was incorporated.  Here the period $\p0$ was set according to Eq.~(\ref{eq-p0}), the generalized Lau condition derived in this work.  As a comparison, in the $4^{\mathrm{th}}$ setup another source grating $G_0^{\ast}$ of period $\p0^{\ast}=12.25 \mu$m was used. This period $\p0^{\ast}$ was set according to Eq.~(\ref{eq-p0*}),  the Lau condition used in\cite{Bopp}.  The aperture-sizes of both source gratings were set to the same as $a=2\mu$m. Intensity fringe formation was simulated numerically as results of wave Fresnel diffraction propagating from the source grating, through the two phase gratings,  and finally to the imaging detector.

The curves in Fig.~\ref{fig-simu-nonEqualP}(a), are the plots of intensity patterns associated with different source settings. In Fig.~\ref{fig-simu-nonEqualP}(a) four curves are presented, they all have the same period of $\pfr = 191.4 \mu$m but different fringe visibility. The blue curve  is the plot for the setup with the point source, while the black curve is the plot for the setup with a 0.35 mm-wide source but without source grating.  Compared to the blue curve, the black curve shows diminishing fringe visibility. This fringe visibility loss is  due to lack of sufficient spatial coherence because of the source width. Incorporating a source grating can help but only if its period is set according to the generalized Lau condition of Eq.~(\ref{eq-p0}). The red and green curves in Fig.~\ref{fig-simu-nonEqualP}(a) correspond to the setups with the same $0.35$mm wide source but with different source grating periods. The red curve is for the source grating setup based on Eq.~(\ref{eq-p0}), that is, the generalized Lau condition. The red curve demonstrates good fringe visibility, which is close to that with the point source. The difference lies in finite aperture size ($a=2 \mu$m) of the source grating. The smaller the aperture size is, the better the fringe visibility. Hence, the red curve clearly validates the generalized Lau condition for determination of source grating period. In stark contrast, the green curve, which is associated with the source grating setup based on Eq.~(\ref{eq-p0*}), depicts as poor fringe visibility as the setup without source grating does. The black and red/green curves demonstrate that incorporating a source grating can restore fringe visibility for setups with finite focal spots, provided that source grating period is set according to the generalized Lau condition of Eq.~(\ref{eq-p0}). The green curve shows clearly that a source grating set with the Lau condition of Eq.~(\ref{eq-p0*}) is unable to restore fringe visibility.  Hence the Lau condition of Eq.~(\ref{eq-p0*}) is not applicable for dual phase grating interferometry. Several other interferometer setups have also been tested, all confirmed the validity of the generalized Lau condition of Eq.~(\ref{eq-p0}) for incorporating source gratings.

In addition, in Fig.~\ref{fig-simu-nonEqualP}(b), (c) and (d) we present the intensity maps obtained when detector pixel size is set to $19.33 \mu$m.  These intensity maps are corresponding to setups associated with the blue, red and green curves in Fig.~\ref{fig-simu-nonEqualP}(a), respectively.

\section{Discussion and conclusions}\label{sec-conclude}

Dual phase grating x-ray interferometry is a promising new technique of grating based x-ray differential phase contrast imaging\cite{Miao, Kagias, Y-W-L-dual-grating-1}. To implement this new X-ray interferometry technique for a setup with an X-ray tube of finite focal spot, an absorbing source grating is required. A source grating serves as an aperture mask to divide the focus spot into an array of mutually incoherent virtual line sources\cite{Pfeiffer, Momose-Huwakara}. In order to attain good fringe visibility, a necessary condition is that the fringe pattern generated by all the line-sources in the source grating are constructively superimposed.  This requirement imposes a stringent design criterion on the period of a source grating in dual phase grating x-ray interferometry. Unfortunately, currently there is confusion in literature about this important design criterion.  As is mentioned earlier, some authors argue that the well-known Lau-condition in Talbot-Lau interferometry can be literally transferred to dual phase grating interferometry\cite{Bopp}. In their reasoning they fail to consider effects on intensity fringes of x-ray wave propagation from the second grating to the imaging detector. Based on this flawed reasoning, they claim that the period of a source grating should be set according Eq.~(\ref{eq-p0*}), the Lau condition for dual phase grating interferometry. Note that the Lau-condition Eq.~(\ref{eq-p0*}) is independent of the distance between the second grating and detector. However, as is shown in section~\ref{sec-results}, a source grating setup based on Eq.~(\ref{eq-p0*}) fails to provide good fringes visibility. 
In this work, we set out to clarify the design criterion on source grating periods in dual phase grating x-ray interferometry. We call this design criterion as the generalized Lau-condition.  In the derivation we noted that it is important to consider the effects on intensity fringe of full wave propagation from the source to the detector. Hence it is incorrect to ignore the effects of wave propagation from second phase grating to the detector.  Although one can derive the generalized Lau-condition from tedious derivation of coherence degree of x-ray illumination\cite{Y-W-L-dual-grating-1, Miao},  but in this work we rather give a simpler and more intuitive derivation of the generalized Lau-condition of Eq.~(\ref{eq-p0}), as is presented in section~\ref{sec-method}. In fact, Eq.~(\ref{eq-p0}) is a necessary condition under which fringe patterns generated by all the line-sources in the source grating are constructively superimposed.  The simulation results presented in section~\ref{sec-results} validated generalized Lau-condition of Eq.~(\ref{eq-p0}) as a design criterion on source grating periods in dual phase grating interferometry.

One remark on Eq.~(\ref{eq-p0}) is on order. In practice one usually has $R_s + R_d \!\gg\! R_g$, hence the pitch $p_0$ of the source grating is much larger than the phase grating pitches $p_1$  and $p_2$. This is a significant advantage of the dual phase grating setup as compared to the inverse geometry setup in Talbot interferometry\cite{Donath-Pfeiffer}. Although both allow to use large detector pixels for intensity fringe detection, but in the inverse geometry the pitch of source grating required to be as small as of few micrometers\cite{Donath-Pfeiffer}. Such narrow-pitch absorbing source grating of high-aspect ratios is hard to fabricate. With dual phase grating setups, Eq.~(\ref{eq-p0}) shows that the pitches of the source grating can be as large as few tens micrometers, thereby such source gratings are easier to fabricate. 

In conclusion, in this work we clarify the issue of what is the condition on source grating period for good fringe visibility. In literature some authors claim that the Lau-condition in Talbot-Lau interferometry can be literally transferred to dual phase grating interferometry. In this work we show that this statement in literature is incorrect and the Lau-condition is not applicable to dual phase grating interferometry. Instead, through an intuitive geometrical analysis of fringe formation, we derived a new generalized Lau-condition for dual phase grating interferometry. The generalized Lau-condition provides a useful design tool for implementation of dual phase grating interferometry.

\section*{Funding}
National Institutes of Health (NIH) (1R01CA193378). 

\end{document}